\documentclass[10pt,letterpaper,twocolumn]{article} 

\usepackage{ol2}
\usepackage[draft]{hyperref}
\usepackage{amsmath}

\begin{document}

\twocolumn[ 

\title{Equivalence principle and quantum mechanics: quantum simulation with entangled photons}


\author{Stefano Longhi}

\address{Dipartimento di Fisica, Politecnico di Milano and Istituto di Fotonica e Nanotecnologie del Consiglio Nazionale delle Ricerche, Piazza L. da Vinci 32, I-20133 Milano, Italy (stefano.longhi@polimi.it)}

\begin{abstract}Einstein`s equivalence principle states the complete physical equivalence of a gravitational field and corresponding inertial field in an accelerated reference frame.  However, to what extent the equivalence principle remains valid in non-relativistic quantum mechanics is a controversial issue. To avoid violation of the equivalence principle, Bargmann`s superselection rule forbids a coherent superposition of states with different masses. Here we suggest a quantum simulation of non-relativistic Schr\"odinger particle dynamics in non-inertial reference frames, which is based on propagation of polarization-entangled photon pairs in curved and birefringent optical waveguides and  Hong-Ou-Mandel quantum interference measurement. The photonic simulator can emulate superposition of mass states, which would lead to violation of the EP. 
\end{abstract}

 ] 

{\it Introduction.} Einstein`s equivalence principle (EP) plays a central role in the theory of gravity. The implications and validity of the EP in quantum physics are a great challenge of current research. There are several forms of the EP \cite{r8}. The `weak` form of the EP states the  universality of free fall, i.e. the equivalence between inertial mass and gravitational mass, whereas the `strong` form of the EP states that 
all effects of a uniform gravitational field are identical to the effects of a uniform acceleration of the coordinate system \cite{r1}.
While the validity of the weak EP for quantum particles has been tested in several experiments (see e.g. \cite{r14}), to what extent the strong form of EP in classical mechanics has an analog in non-relativistic quantum mechanics
remains controversial \cite{r2,r3,r4,r5,r6,r8,r9,r10}. 
Earlier experiments of quantum interference with neutrons confirmed the validity of the strong form of EP for non-relativistic quantum particles \cite{r11,r12,r13}.
 However, controversies arise if the particle is allowed to be in superposition states of different
masses. Covariance of the Schr\"odinger equation under extended Galileian boosts requires a phase change of the wave function, which is dependent on the particle mass and related to the particle proper time \cite{r6,r14,r15}. For particle states with definite mass, such a phase term does not pose any difficulty, however if a coherent superposition of states of different masses is allowed, this would lead to different evolutions in different reference frames  \cite{r6,r14}. To ensure the compatibility of non-relativistic quantum mechanics with EP, one generally invokes a superselection rule \cite{r16}, originally proposed by Bargmann \cite{r17}, which prevents a quantum particle from being found in superposition states of different masses. Bargmann`s superselection rule is generally considered an expedient \cite{r6}, and ways to overcome it have been suggested by some authors \cite{r6,r10,r18,r19}. The main trouble is that, while superposition of different mass states may be possible for subatomic particles (e.g. for neutrino \cite{r19bis}), in non-relativistic quantum mechanics mass is a kinematical parameter and superposition states are conceptually unclear states \cite{r5,r19}: Bargmann`s superselection rule is thus the result of a Gedankenexperiment.\\ 
Recently, emulation of gravity effects that still elude observation have 
been suggested using optical and electromagnetic setups \cite{r20,r21,r22,r23,r24,r25,r26,r27,r28,r29,r29bis}. These include rotating black holes \cite{r20,r29bis}, event horizon \cite{r21},  Hawking radiation \cite{r25}, quantum gravity \cite{r26,r29}, and self-gravitation effects \cite{r27,r28}. Quantum optics experiments, using single photons or entangled-photon pairs, have been also proposed to probe quantum mechanics in curved space-time and to test gravity-induced decoherence effects \cite{uffa1,uffa2}.\\ 
In this Letter I propose a quantum optical simulator of non-relativistic dynamics of a quantum particle in non-inertial reference frames that enables to emulate coherent superposition of different mass states. The quantum simulator consists of spatial propagation of a polarization-entangled photon state (a Bell state) in curved and weakly-birefringent optical waveguides. Interference of mass superposition states is emulated by two-photon quantum interference on a beam splitter \cite{r40}, a kind of measurement which is routinely done in quantum optics.\par
{\it  Einstein's equivalence principle in non-relativistic quantum mechanics.} Let us first briefly recall  the covariance property of the Schr\"odinger equation under extended  Galilean transformations \cite{r3,r6,r15,r30}. In an inertial reference frame $S$, with space and time variables $x$ and $t$, the dynamics of a non-relativistic spinless quantum particle of mass $m$ subjected to a scalar potential $V(x,t)$ is governed by the Schr\"odinger equation
\begin{equation}
i \hbar \frac{\partial \psi}{\partial t}=-\frac{\hbar^2}{2m} \frac{\partial^2 \psi}{\partial x^2} +V(x,t)\psi
\end{equation}
for the particle wave function $\psi=\psi(x,t)$. Let us consider a coordinate transformation from the inertial frame $S$
to an accelerated one $S`$ via the extended Galilean transformation \cite{r6,r15,r30}
\begin{equation}
x`= x - \xi(t), \; \; t`= t. 
\end{equation}
where $\xi(t)$ describes the rigid translational motion of $S`$ with respect to $S$. After setting
\begin{equation}
\psi(x`,t`)=\psi`(x`,t`) \exp \left[ i \frac{m \dot{\xi} x`}{\hbar} +i \varphi_m(t`) \right]
\end{equation}
where the dot indicates the derivative with respect to time and 
\begin{equation}
\varphi_m(t`) \equiv \frac{m}{2 \hbar} \int_0^{t`} d \rho \dot{\xi}^2(\rho),
\end{equation}
from Eqs.(1-3) one readily obtains
\begin{equation}
i \hbar \frac{\partial \psi`}{\partial t`}=-\frac{\hbar^2}{2m} \frac{\partial^2 \psi`}{\partial x`^2} +V`(x`,t`) \psi`
\end{equation}
where we have set
\begin{equation}
V`(x`,t`) \equiv V(x,t)+m \ddot{\xi} x`.
\end{equation}
Equations (5) and (6) show that the non-inertial observer $S`$ describes the particle motion using the same form of Schr\"odinger equation as $S$, however the potential
$V`(x`,t`)$ acting on the particle differs from $V$  because of the additional uniform gravitational field  $m \ddot{\xi} x`$, which is exactly what one expects from
 the EP of classical mechanics. However, to ensure covariance of the Schr\"odinger equation the wave functions $\psi$ and $\psi`$ in the two reference frames differ each other for a mass-dependent phase term [Eqs.(3) and (4)], which could eventually cause mass-dependent diffraction and interference effects \cite{r6,r30}. Such a mass-dependent phase term indicates that, strictly speaking, non-relativistic quantum mechanics under extended Galilean transformations does not fulfill the strong version of the EP.\\
As an example, let us assume $\xi(t)=\dot{\xi}(t)=0$ for $t \leq 0$ and $t \geq T$,
so that $S`$ takes a closed circuit and coincides with system $S$  at times $t \leq 0$ and $t \geq T$, while in the interval $(0,T)$ $S`$ is accelerating \cite{r6}. According to Eq.(3) one has $\psi`(x,T)=\psi(x,T) \exp (i \varphi_m)$ where $\varphi_m= (m/ 2 \hbar) \int_0^T dt \dot{\xi}^2(t)$.  The phase shift $\varphi_m$ has an interesting physical interpretation, namely $\varphi_m=mc^2 \Delta \tau / \hbar$, where $\Delta \tau$ is the difference of proper times $\tau=t$ and $\tau`=\int_0^t d \rho (1-\dot{\xi}^2/c^2)^{1/2}$ between the two coordinate systems $S$ and $S`$ in the non-relativistic limit $|\dot{\xi}| \ll c$ \cite{r6}. In other words, the phase shift $\varphi_m$ between wave functions $\psi$ and $\psi`$ in the two reference frames comes from the fact that the clock in $S`$ runs at a different rate than that in $S$, so as when one has arrived back in $S$ at time $T$ less time has been passed in the system $S`$ than in the system $S$ (like in the twin paradox of special relativity). Let us now assume in $S$ a rigidly-moving (for example oscillating) potential $V(x,t)=W(x-\xi(t))$, where $W(x)$ is a potential well that sustains one (or more) bound states, and let as assume that at time $t=0$ the particle state is prepared in the ground state $u(x)$ of the potential well, i.e. $\psi(x,0)=\psi`(x,0)=u(x)$; see Fig.1. In system $S`=(x`=x-\xi(t),t`=t)$, the potential well appears to be at rest and the wave function $\psi`$ evolves according to Eq.(5) with the potential $V`(x`,t`)=W(x`)+m \ddot{\xi} x`$. 
\begin{figure}[htb]
\centerline{\includegraphics[width=8.4cm]{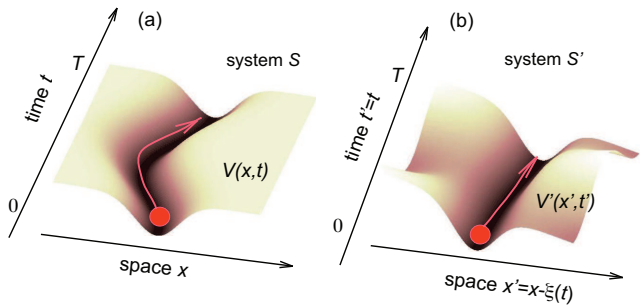}} \caption{ \small
(Color online) (a) Motion of a quantum particle subjected to an accelerated (oscillating) potential well $V(x,t)=W(x-\xi(t))$ as observed in the inertial reference frame $S=(x,t)$. (b) The same motion as observed in the accelerating (non-inertial) reference frame $S`=(x`=x-\xi(t),t`)$. The transformed potential is $V`(x`,t`)=W(x`)+m \ddot{\xi}(t)x`$. For a slowly-moving potential well the effect of the gravitational field $m \ddot{\xi}(t)x`$ on particle motion in system $S`$ can be neglected. Different proper times in $S$ and $S`$ result in a non vanishing phase difference $\varphi_m$ of wave functions.}
\end{figure} 
The effect of the time-dependent gravitational potential $m \ddot{\xi} x`$ entering in $V`$ is to induce transitions from the ground state to upper excited states or to the continuum of states of the potential well $W$. The transition probability can be calculated using standard time-dependent perturbation theory. However, in principle one can consider a sufficiently slow motion $\xi(t)$ so that transitions to other states are negligible. Since $\dot{\xi}(0)=\dot{\xi}(T)=0$, in system $S`$ one readily obtains $\psi`(x,T) \simeq u(x) \exp(-i E_m T/ \hbar)$, where $E_m$ is the ground state energy of the potential well: in other words, the effect of the inertial force on particle motion in $S`$ can be neglected, the particle remains in the ground state $u(x)$ and the wave function just acquires the phase shift $-E_m T/ \hbar$ of stationary state. On the other hand, in system $S$ one has $\psi(x,T)=\psi`(x,T) \exp(i \varphi_m)$. In ordinary non-relativistic quantum mechanics the particle mass $m$ is a kinematical parameter in the Schr\"odinger equation. However, one can imagine a {\em Gedankenexperiment} where the initial quantum state is a coherent superposition of states with different masses $\psi_{m_1}$ and $\psi_{m_2}$ \cite{r6}, i.e. $\psi=\psi_{m_1}+\psi_{m_2}$. Assuming that  $\psi_{m_1}$ and $\psi_{m_2}$ independently evolve according to the Schr\"odinger equation (with either $m=m_1$ or $m=m_2$), system $S`$ will induce a different phase shift between the two mass state components, relative to the first system $S$, which could be detectable in an interference experiment. This would lead to violation of the EP.  Bargmann`s superselection rule  $\Delta m=m_2-m_1=0$ is thus viewed as a necessary requirement to avoid EP violation \cite{r16,r17}. However, as pointed out by several authors the above argument rises some conceptual problems   \cite{r5,r6,r10,r18,r19}, and extended theories beyond ordinary quantum mechanics would be in order to ensure compatibility between the EP and quantum mechanics without requiring a mass superselection rule (for example considering mass and proper time as conjugate variables). The main caveat underling derivation of Bargmann`s superselection rule is that it imposes restrictions on a kinematical level for states that, on the dynamical level, we would not know exactly how to handle \cite{r5,r19}: if mass enters in the Schr\"odinger equation as a kinematical (rather than dynamical) variable, how can one construct a coherent superposition of mass states in a Gedanken experiment? And how should they evolve?\par
{\it Quantum simulation with entangled photons.} Here we suggest a quantum simulation of the Gedankenexperiment mentioned above, which enables to emulate the coherent superposition of states with different masses and to measure the phase shift $\varphi_m$  related to the different clock rates in the two reference frames. The quantum simulation is based on the similarity between spatial propagation of  monochromatic light waves in weakly-curved optical waveguides and the quantum dynamics of a non-relativistic particle in a non-inertial reference frame $S`$ \cite{r31}. In such an analogy, the local curvature of the optical waveguide  axis emulates the instantaneous acceleration of system $S`$. In other words, optical waveguide axis bending introduces an equivalent gravitational potential for photons, which has found several applications, for example in dynamical localization of light \cite{r32,r33}, control of photon tunneling \cite{r34} and realization of synthetic gauge fields for photons \cite{r35,r36} (for a recent review see \cite{r37}). In the scalar and paraxial approximations, propagation of monochromatic light
waves at wavelength $\lambda$ in a weakly-guiding dielectric waveguide, whose axis is 
curved along the paraxial propagation direction $z$ [Fig.2(a)], is described by the optical Schr\"odinger equation for the electric field envelope $\psi(x,y,z)$ \cite{r31}
\begin{equation}
i (1/k) \frac{\partial \psi}{\partial z}=-\frac{1}{2 n_s k^2} \frac{\partial^2 \psi}{\partial x^2} +W(x-\xi(z)) \psi
\end{equation}
where $k=2 \pi / \lambda$, $W (x) = [ n^{2}_{s}-n^2(x)]/(2ns) \simeq n_s- n(x)$ is the optical potential, $n(x)$ is the refractive index profile of the waveguide, $n_s$ is the cladding refractive index, and $\xi(z)$ is the axis bending profile. Clearly Eq.(7) is analogous to the Schr\"odinger equation (1) in an inertial reference frame $S$ provided that the formal substitutions $t \rightarrow z$, $ \hbar \rightarrow 1/k$, $m \rightarrow n_s$ and $V \rightarrow W$ are made.  The optical potential $W(x)$ is the analog of the potential well discussed above and $u(x)$ is the profile of the guided optical mode. In the waveguide reference frame $S`$, $x`=x-\xi(z)$ and $z`=z$, the bending is replaced by a refractive index gradient (the gravitational field), according to the EP. As discussed above, for a weakly bent waveguide axis the gradient potential does not induce transitions and can be effectively neglected. The phase difference $\varphi_m$ entering in Eq.(3) has a simple geometrical interpretation in our optical analogue: it corresponds to the different optical path of the light wave experienced when propagating in a straight and in a bent waveguide. In fact, while in $S`$ the waveguide appears straight and the longitudinal propagation phase associated to the optical path $n_sz$ is $kn_sz$, in $S$ the waveguide is bent and the propagation phase term of the wave is $kn_s s$, where $s=\int^z d\rho \sqrt{1+ \dot{\xi}^2}$ is the curvilinear abscissa and $\dot \xi \equiv d \xi / d \rho$. In the paraxial regime $|\dot \xi | \ll 1$, one has $s \simeq z+(1/2) \int^z d \rho \dot{\xi^2}$ and hence $k n_s s \simeq kn_s z+\varphi_m(z)$, where $\varphi_m(z)=(kn_s/2)\int^z d \rho \dot{\xi}^2$ is precisely the additional phase term (4).
Note also that in the quantum-optical analogue the particle mass $m$ is replaced by the refractive index $n_s$. If the waveguide shows a weak birefringence, as it happens for instance in waveguides currently used in integrated quantum photonics circuits \cite{r38}, the refractive index $n_s$ takes slightly different values, $n_{s,H}$ and $n_{s,V}$, depending on the polarization state (horizontal H or vertical V) of the excitation wave \cite{r38,r39}.  The paraxial equation (7) can be used to describe propagation of quantum light as well \cite{referee1}. If the waveguide is excited by non-classical light in a polarization entangled state, a coherent superposition of states with different masses and violation of the EP can be effectively emulated. To this aim, let us consider the optical setup shown in Fig.2(b). It consists of two equal, widely-spaced and weakly-birefringent waveguides, one straight (waveguide 1) and the other one weakly curved for a length $T$ (waveguide 2) with a bending profile $\xi(z)$ satisfying the conditions $\xi(z)=\dot{\xi}(z)=0$ for $z=0$ and $z=T$. The waveguide system is excited at input plane by a polarization-entangled 
two-photon state (a $ | \phi^+ \rangle$ Bell state), which is routinely generated in type-II frequency down-conversion \cite{referee2}
 \begin{figure}[htb]
\centerline{\includegraphics[width=8.4cm]{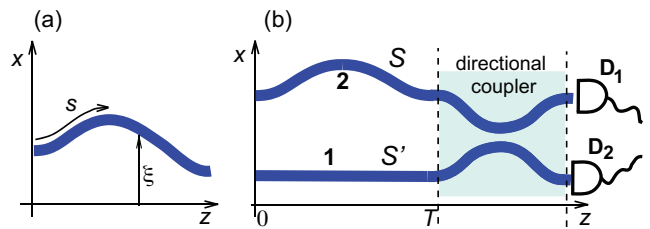}} \caption{ \small
(Color online) (a) Schematic of an optical waveguide with a bent axis. $\xi=\xi(z)$ is the bending profile; $s$ is the arc length along the curved waveguide axis. (b) Optical setup for the quantum simulation of EP violation. Two weakly-birefringent and widely-spaced optical waveguides 1 and 2 are excited at the input plane $z=0$ by a two-photon polarization entangled state. The optical axis of waveguide 2 is weakly bent, while the axis of waveguide 1 is straight. Photon propagation in the two guides emulates the dynamical behavior of a quantum particle as observed in reference frames $S$ (waveguide 2) and $S`$ (waveguide 1). At the output plane $z=T$ photons are mixed in a balanced directional coupler, which acts as a $50\%$ beam splitter. Two-photon coincidence at photodetectors $D_1$ and $D_2$ is measured.}
\end{figure} 
\begin{equation}
| \phi \rangle_{in} = \frac{1}{\sqrt{2}} \left( \hat{a}^{\dag (H)}_1 \hat{a}^{\dag (V)}_2 |0 \rangle + \hat{a}^{\dag (V)}_1 \hat{a}^{\dag (H)}_2 |0 \rangle \right)
\end{equation}
where $\hat{a}^{\dag (H,V)}_i$ is the creation operator of photons in waveguide $i$ ($i=1,2$) with either horizontal ($H$) or vertical ($V$) polarization and $|0 \rangle$ is the vacuum state. Each photon can propagate in either one of the two waveguides: propagation in waveguide 1 emulates the observer $S`$,  while propagation in waveguide 2 emulates the observer $S$. The EP would predict the equivalence of propagation in the two waveguides: this means that any interference measurement at the output plane $z=T$ should be independent of the bending profile $\xi(z)$ of waveguide 2. The photon state at the output plane is simply given by
\begin{equation}
| \phi \rangle_{out} = \frac{1}{\sqrt{2}} \left( \exp(i \theta_1) \hat{a}^{\dag (H)}_{1} \hat{a}^{\dag (V)}_{2} |0 \rangle + \exp(i \theta_2) \hat{a}^{\dag (V)}_{1} \hat{a}^{\dag (H)}_{2} |0 \rangle \right)
\end{equation}
where $\theta_1$ and $\theta_2$ are the longitudinal phase shifts arising from wave propagation in the two waveguides with different optical paths, namely
\begin{equation}
\theta_1 \equiv k(n_{s,H}T+n_{s,V}s) \; , \; \; \theta_2 \equiv  k(n_{s,V}T+n_{s,H}s)
\end{equation}
and $s\simeq T+(1/2)\int_0^T d \rho \dot{\xi}^2$ is the arc length of the curved waveguide 2, from $z=0$ to $z=T$. 
 We then perform two-photon (Hong-Ou-Mandel) quantum interference \cite{r40} by combining the two photons via a $50\%$ beam splitter, which is a standard setup in quantum optics \cite{referee2,referee3,referee4}. The beam splitter is realized by bending the waveguides close to each other realizing a balanced optical directional coupler \cite{r38}. Indicating by $\hat{b}^{\dag (H,V)}_i$ the creation operator of photons, with either $H$ or $V$ polarization, at the output waveguide port $i$ ($i=1,2$) of the coupler, the photon state after propagation across the directional coupler is obtained from Eq.(9) with the replacements
\begin{equation}
\hat{a}^{\dag (H,V)}_{1} \rightarrow \frac{\hat{b}^{\dag (H,V)}_{1}+\hat{b}^{\dag (H,V)}_{2}}{\sqrt{2}} \; ,\; \hat{a}^{\dag (H,V)}_{2} \rightarrow \frac{\hat{b}^{\dag (H,V)}_{1}-\hat{b}^{\dag (H,V)}_{2}}{\sqrt{2}}
\end{equation}
Placing two detectors at the output coupler ports [Fig.2(b)], the coincidence probability to measure one photon in each port, regardless of the polarization state of the photons, is given by $P^{(1,1)}= \sin^2 (\Delta \varphi /2)$
where $\Delta \varphi= \theta_2-\theta_1$, i.e.
\begin{equation}
\Delta \varphi =k(n_{s,H}-n_{s,V})(s-T)=\frac{k}{2} (n_{s,H}-n_{s,V}) \int_0^T d \rho \dot{\xi}^2.
\end{equation}
Interestingly, $\Delta \varphi$ is precisely the difference between phases $\varphi_{m_1}$ and $\varphi_{m_2}$ associated to the different masses $m_1=n_{s,H}$ and $m_2=n_{s,V}$. Clearly $\Delta \varphi$  requires both the extra path length of curved waveguide and modal birefringence: $\Delta \varphi$ vanishes for vanishing modal birefringence or for balanced lengths of the two waveguides. A two-photon quantum interference measurement is able to detect such a phase difference: for $\Delta m=0$, i.e. for $n_{s,V}=n_{s,H}$,  two photons bounce together and the two-photon coincidence probability at detectors $D_1$ and $D_2$ vanishes. However, for $\Delta m \neq 0$  a non-vanishing coincidence probability (photon antibunching) can be observed. In particular, for $\Delta \varphi = \pi$ photon antibunching is observed with $P^{(1,1)}=1$, corresponding to switching between the two maximally-entangled (Bell) states $| \phi^+ \rangle$ and $| \phi^- \rangle$. For example, let as consider a typical birefringence $n_{s,H}-n_{s,V} \simeq 1 \times 10^{-4}$ at $\lambda \simeq 815 \; {\rm nm}$ found in femtosecond-laser-written waveguides \cite{r38,r39}. The largest photon antibunching ($\Delta \varphi= \pi$) is observed for a length mismatch $s-T \simeq 4\; {\rm mm}$. For a sinusoidally-bent waveguide $\xi(z)=A[1-\cos ( 2 \pi z/T)]$  of length $T=8$ cm, a transverse shift amplitude $A \simeq 5.7 \; {\rm mm}$ is required. By halving the amplitude $A$, $P^{(1,1)}$ is reduced to $ \sin^2 ( \pi/8) \simeq 0.15$. \par
{\em Conclusions.} The strong version of the EP establishes that the effects of a uniform gravitational field can not be distinguished from those of a uniform acceleration of the coordinate system. Is the EP compatible with quantum mechanics ? The short answer is that it is. However, when dealing with superposition of states with different masses the  EP seems to be in trouble. Gedanken experiments, showing the incompatibility between EP and quantum mechanics, would involve superposition of mass states \cite{r6}. But this poses some conceptual and practical problems, since we do not know exactly how to handle the mass in quantum mechanics and how to construct mass superposition states \cite{r5,r18,r19}. Here I suggested a quantum simulation of EP in non-relativistic quantum mechanics on an integrated photonic chip, which is based on the propagation of entangled photon pairs in curved and weakly-birefringent optical waveguides. The optical setting can emulate  coherent superposition of states with different mass and reveal interferometrically mass-dependent phase effects. Our results are expected to inspire further studies aimed at testing the interplay of gravitation and the quantum-mechanical principle of linear superposition. For example, since Hong-Ou-Mandel quantum interference can occur for atoms \cite{r41} and inertial forces can be realized by accelerating optical lattices, cold atoms could provide another platform for a quantum simulation. As compared to photons, exploiting internal degrees of freedom of the atoms one might be able to emulate superposition states with more than two different effective masses or even models where mass is an effective dynamical variable.

\newpage


 {\bf References with full titles}\\
 \\
 \noindent
 1. E. Okon and C. Callender, {\it Does quantum mechanics clash with the equivalence principle-and does it matter?}, Eur. J. Philos. Sci. {\bf 1}, 133
(2011).\\
 2. A. Einstein, {\it On the relativity principle and the conclusions drawn from it}, Jahr. Radioaktivitit\"at Elektron. 4, 411 (1907).\\
 3. G. Rosi, G. D'Amico, L. Cacciapuoti, F. Sorrentino, M. Prevedelli, M. Zych, C. Brukner, and G. M. Tino, {\it Quantum test of the equivalence principle for atoms in superpositions of internal energy eigenstates}, Nat. Commun. {\bf 8}, 15529 (2017).\\
 4. D. Greenberger, {\it The Role of Equivalence in Quantum Mechanics}, Ann. Phys. {\bf 47}, 116 (1968).\\
 5. C. J. Eliezer and P. G. Leach , {\it The equivalence principle and quantum mechanics}, Am. J. Phys. {\bf 45}, 1218 (1977).\\
 6.  C. L\"ammerzahl , {\it On the equivalence principle in quantum theory}, Gen. Relativ. Gravitation {\bf 28}, 1043 (1996).\\ 
 7. D. Giulini, {\it On Galilei invariance in quantum mechanics and the Bargmann superselection rule}, Ann. Phys. {\bf 249}, 222 (1996).\\
 8. D.M. Greenberger, {\it Inadequacy of the usual galilean transformation in quantum mechanics}, Phys. Rev. Lett. {\bf 87}, 100405 (2001).\\
 9. H. Hernandez-Coronado, {\it From Bargmann's superselection rule to quantum Newtonian spacetime}, Found. Phys. {\bf 42}, 1350 (2012).\\
 10. S. T. Pereira and R. M. Angelo, {\it Galilei covariance and Einstein's equivalence principle in quantum reference frames}, Phys. Rev. A {\bf 91}, 022107 (2015).\\
 11. R. Colella, A. W. Overhauser, and S. A. Werner, {\it Observation of gravitationally induced quantum interference}, Phys. Rev. Lett. {\bf 34}, 1472 (1975).\\
 12. D. M. Greenberg and A. W. Overhauser , {\it Coherence effects in neutron diffraction and gravity experiments}, Rev. Mod. Phys. {\bf 51}, 43 (1979).\\
 13. U. Bonse and T. Wroblewski, {\it Measurement of Neutron Quantum Interference in Noninertial Frames}, Phys. Rev. Lett. {\bf 51}, 1401 (1983).\\
 14. G. Rosen, {\it Galilean Invariance and the General Covariance of Nonrelativistic Laws}, Am. J. Phys. {\bf 40}, 683 (1972).\\
 15.  L. E. Ballentine, {\it Quantum Mechanics: A Modern Development} (World Scientific, Singapore, 1998).\\
16. U. Bargmann, {\it On Unitary Ray Representations of Continuous Groups}, Ann. Math. {\bf 59}, 1 (1954).\\
17. D. M. Greenberger, {\it Wavepackets for particles of indefinite mass}, J. Math. Phys. {\bf 15}, 406 (1974).\\
18. H. Hernandez-Coronado and E. Okon, {\it Quantum equivalence principle without mass superselection}, Phys. Lett. A {\bf 377}, 2293 (2013).\\
19. D. V. Ahluwalia and C. Burgard, {\it Interplay of gravitation and linear superposition of different mass eigenstates}, Phys. Rev. D {\bf 57}, 4724 (1998).\\
20. I.I. Smolyaninov, {\it Surface plasmon toy model of a rotating black hole}, New J. Phys. {\bf 5}, 147 (2003).\\
21. T.G. Philbin, C. Kuklewicz, S. Robertson, S. Hill, F. K\"onig, and U. Leonhardt, {\it Fiber-Optical Analog of the Event Horizon},
Science {\bf 319}, 1367 (2006).\\
22. D.A. Genov,  S. Zhang and X. Zhang, {\it Mimicking celestial mechanics in metamaterials}, Nat. Phys. {\bf 5}, 687 (2009).\\
23. V.H. Schultheiss, S. Batz, A. Szameit, F. Dreisow, S. Nolte, A. T\"unnermann, S. Longhi, and U. Peschel, {\it Optics in Curved Space}, Phys. Rev. Lett. {\bf 105}, 143901 (2010).\\
24. U. Leonhardt, C. Maia, and R. Sch\"utzhold, {\it Focus on classical and quantum analogues for gravitational phenomena and related effects}, New J. Phys. {\bf 14}, 105032 (2012).\\
25. F. Belgiorno, S.L. Cacciatori, M. Clerici, V. Gorini, G. Ortenzi, L. Rizzi, E. Rubino, V.G. Sala, and D. Faccio, {\it Hawking radiation from ultrashort laser pulse filaments}, Phys. Rev. Lett. {\bf 105}, 203901 (2010).\\
26. C. Conti, {\it Quantum gravity simulation by non-paraxial nonlinear optics}, Phys. Rev. A {\bf 89}, 061801(R) (2014).\\
27. R. Bekenstein, R. Schley, M. Mutzafi, C. Rotschild, and M. Segev, {\em Optical simulations of gravitational effects in the
Newton-Schr\"odinger system}, Nat. Phys. {\bf 11}, 872 (2015).\\
28. T. Roger, C. Maitland, K. Wilson, N. Westerberg, D. Vocke, E.M. Wright,
and D. Faccio, {\it Optical analogues of the Newton-Schr\"odinger equation and boson star evolution}, Nat. Commun. {\bf 7}, 13492 (2016).\\
29. M. C. Braidotti, Z.H. Musslimani, and C. Conti, {\it Generalized Uncertainty Principle and Analogue of Quantum Gravity in Optics}, Physica D {\bf 338}, 34 (2017).\\
30. D. Vocke, C. Maitland, A. Prain, F. Biancalana, F. Marino, E.M. Wright, and D. Faccio, {\it Rotating black hole geometries in a two-dimensional photon superfluid}, arXiv:1709.04293 (2017).\\
31. M. Zych, F. Costa, I. Pikovski, T.C. Ralph, and C. Brukner, {\it General relativistic effects in quantum interference of photons}, Class. Quantum Grav. {\bf 29}, 224010 (2012).\\
32. T.C. Ralph, G.J. Milburn, and T. Downes, {\it Gravitationally induced decoherence of optical entanglement}. Preprint at http://arxiv.org/abs/quant-ph/0609139
(2006).\\
33. C.K. Hong, Z.Y. Ou, and L. Mandel, {\it Measurement of subpicosecond time intervals between two photons by interference}, Phys. Rev. Lett. {\bf 59},  2044 (1987).\\ 
34. D.M. Greenberger, {\it Some remarks on the extended Galilean transformation}, Am. J. Phys. {\bf 47}, 35 (1979).\\
35. S. Longhi, {\it Quantum-optical analogies using photonic structures},  Laser \& Photon. Rev. {\bf 3}, 243 (2009).\\
36. S. Longhi, M. Marangoni, M. Lobino, R. Ramponi, P. Laporta, E. Cianci, and V. Foglietti, {\it Observation of Dynamic Localization in Periodically-Curved Waveguide Arrays}, Phys. Rev. Lett. {\bf 96}, 243901 (2006).\\
37. A. Szameit, I.L. Garanovich, M. Heinrich, A.A. Sukhorukov, F. Dreisow, T. Pertsch, S. Nolte, A. T\"unnermann, and Y.S. Kivshar, {\it Polychromatic dynamic localization in curved photonic lattices}, Nat. Phys. {\bf 5}, 271 (2009).\\
38. G. Della Valle, M. Ornigotti, E. Cianci, V. Foglietti, P. Laporta, and S. Longhi, {\it Visualization of Coherent Destruction of Tunneling in an Optical Double-Well System}, Phys. Rev. Lett. {\bf 98}, 263601 (2007).\\
39. M.C Rechtsman, J.M. Zeuner, Y. Plotnik, Y. Lumer, D. Podolsky, F. Dreisow, S. Nolte, M. Segev, and A. Szameit, {\it Photonic Floquet topological insulators},  Nature {\bf 496}, 196 (2013).\\
40. S. Longhi, {\it Effective magnetic fields for photons in waveguide and coupled resonator lattices}, Opt. Lett. {\bf 38}, 3570 (2013).\\
41. I.L. Garanovich, S. Longhi, A.A. Sukhorukov, and Y.S. Kivshar, {\it Light propagation and localization in modulated photonic lattices and waveguides}, Phys. Rep. {\bf 518}, 1 (2012).\\
42 T. Meany, M. Gr\"afe, R. Heilmann, A. Perez-Leija, S. Gross, M.J. Steel, M.J. Withford, and A. Szameit, {\it Laser written circuits for quantum photonics}, Laser \& Photon. Rev. {\bf 9}, 363 (2015).\\
43. L.A. Fernandes, J.R. Grenier, P.R. Herman, J.S. Aitchison, and P.V.S. Marques, {\it Stress induced birefringence tuning in femtosecond laser fabricated waveguides in fused silica}, Opt. Express {\bf 20}, 24103 (2012).\\ 
44. S. Longhi, {\it Optical Bloch Oscillations and Zener Tunneling with Nonclassical Light}, Phys. Rev. Lett. {\bf 101}, 193902 (2008).\\
45. P.G. Kwiat, K. Mattle, H. Weinfurter, A. Zeilinger, A.V. Sergienko, and Y. Shih, {\it New High-Intensity Source of Polarization-Entangled Photon Pairs}, Phys. Rev. Lett. {\bf 75}, 4337 (1995).\\
46. R.-B. Jin, R. Shimizu, K. Wakui, M. Fujiwara, T. Yamashita, S. Miki, H. Terai, Z. Wang, and M. Sasaki, {\it Pulsed Sagnac polarization-entangled
photon source with a PPKTP crystal at telecom wavelength}, Opt. Express {\bf 22}, 11498 (2014).\\
47. Y.-H. Li, Z.-Y. Zhou, Z.-H. Xu, L.-X. Xu, B.-S. Shi, and G.-C. Guo, {\it Multiplexed entangled photon-pair sources for all-fiber quantum networks}, Phys. Rev. A {\bf 94}, 043810 (2016).\\
48. R. Lopes, A. Imanaliev, A. Aspect, M. Cheneau, D. Boiron, and C.I. Westbrook. {\it Atomic Hong-Ou-Mandel experiment}, Nature {\bf 520}, 66 (2015).\\

\begin{thebibliography}{99}




\bibitem{r8}
 E. Okon and C. Callender, Eur. J. Philos. Sci. {\bf 1}, 133 (2011).
\bibitem{r1} 
A. Einstein,  Jahr. Radioaktivitit\"at Elektron. 4, 411 (1907).
\bibitem{r14}
G. Rosi, G. D'Amico, L. Cacciapuoti, F. Sorrentino, M. Prevedelli, M. Zych, C. Brukner, and G. M. Tino, Nat. Commun. {\bf 8}, 15529 (2017).
\bibitem{r2}
D. Greenberger, Ann. Phys. {\bf 47}, 116 (1968).
\bibitem{r3}
 C. J. Eliezer and P. G. Leach, Am. J. Phys. {\bf 45}, 1218 (1977).
\bibitem{r4} 
 C. L\"ammerzahl , Gen. Relativ. Gravitation {\bf 28}, 1043 (1996).
\bibitem{r5}
D. Giulini,  Ann. Phys. {\bf 249}, 222 (1996).
\bibitem{r6}
D.M. Greenberger, Phys. Rev. Lett. {\bf 87}, 100405 (2001).
 \bibitem{r9}
  H. Hernandez-Coronado, Found. Phys. {\bf 42}, 1350 (2012).
 \bibitem{r10}
 S. T. Pereira and R. M. Angelo, Phys. Rev. A {\bf 91}, 022107 (2015).
\bibitem{r11}
R. Colella, A. W. Overhauser, and S. A. Werner, Phys. Rev. Lett. {\bf 34}, 1472 (1975).
\bibitem{r12}
D.M. Greenberg and A. W. Overhauser, Rev. Mod. Phys. {\bf 51}, 43 (1979).
\bibitem{r13}
U. Bonse and T. Wroblewski, Phys. Rev. Lett. {\bf 51}, 1401 (1983).
 \bibitem{r15}
G. Rosen,  Am. J. Phys. {\bf 40}, 683 (1972).
 \bibitem{r16}
L. E. Ballentine, {\it Quantum Mechanics: A Modern Development} (World Scientific, Singapore, 1998).
\bibitem{r17}
U. Bargmann, Ann. Math. {\bf 59}, 1 (1954).
\bibitem{r18}
D.M. Greenberger, J. Math. Phys. {\bf 15}, 406 (1974).
\bibitem{r19}
H. Hernandez-Coronado and E. Okon, Phys. Lett. A {\bf 377}, 2293 (2013).
\bibitem{r19bis}
D. V. Ahluwalia and C. Burgard, Phys. Rev. D {\bf 57}, 4724 (1998).
\bibitem{r20}
I.I. Smolyaninov, New J. Phys. {\bf 5}, 147 (2003).
\bibitem{r21}
T.G. Philbin, C. Kuklewicz, S. Robertson, S. Hill, F. K\"onig, and U. Leonhardt, Science {\bf 319}, 1367 (2006).
\bibitem{r22}
D.A. Genov,  S. Zhang and X. Zhang, Nat. Phys. {\bf 5}, 687 (2009).
\bibitem{r23}
V.H. Schultheiss, S. Batz, A. Szameit, F. Dreisow, S. Nolte, A. T\"unnermann, S. Longhi, and U. Peschel, Phys. Rev. Lett. {\bf 105}, 143901 (2010).
\bibitem{r24}
U. Leonhardt, C. Maia, and R. Sch\"utzhold, New J. Phys. {\bf 14}, 105032 (2012).
\bibitem{r25}
F. Belgiorno, S.L. Cacciatori, M. Clerici, V. Gorini, G. Ortenzi, L. Rizzi, E. Rubino, V.G. Sala, and D. Faccio, Phys. Rev. Lett. {\bf 105}, 203901 (2010)
\bibitem{r26}
C. Conti, Phys. Rev. A {\bf 89}, 061801(R) (2014).
\bibitem{r27}
R. Bekenstein, R. Schley, M. Mutzafi, C. Rotschild, and M. Segev, Nat. Phys. {\bf 11}, 872 (2015).
\bibitem{r28}
 T. Roger, C. Maitland, K. Wilson, N. Westerberg, D. Vocke, E.M. Wright, and D. Faccio, Nat. Commun. {\bf 7}, 13492 (2016).
 \bibitem{r29}
M. C. Braidotti, Z.H. Musslimani, and C. Conti, Physica D {\bf 338}, 34 (2017).
\bibitem{r29bis}
D. Vocke, C. Maitland, A. Prain, F. Biancalana, F. Marino, E.M. Wright, and D. Faccio, arXiv:1709.04293 (2017).
\bibitem{uffa1}
M. Zych, F. Costa, I. Pikovski, T.C. Ralph, and C. Brukner, Class. Quantum Grav. {\bf 29}, 224010 (2012). 
\bibitem{uffa2}
T.C. Ralph, G.J. Milburn, and T. Downes, arxiv: quant-ph/0609139
\bibitem{r40}
 C.K. Hong, Z.Y. Ou, and L. Mandel, Phys. Rev. Lett. {\bf 59},  2044 (1987).
\bibitem{r30}
D.M. Greenberg, Am. J. Phys. {\bf 47}, 35 (1979).
\bibitem{r31}
S. Longhi,  Laser \& Photon. Rev. {\bf 3}, 243 (2009).
\bibitem{r32}
S. Longhi, M. Marangoni, M. Lobino, R. Ramponi, P. Laporta, E. Cianci, and V. Foglietti, Phys. Rev. Lett. {\bf 96}, 243901 (2006).
\bibitem{r33}
A. Szameit, I.L. Garanovich, M. Heinrich, A.A. Sukhorukov, F. Dreisow, T. Pertsch, S. Nolte, A. T\"unnermann, and Y.S. Kivshar, Nat. Phys. {\bf 5}, 271 (2009).
\bibitem{r34}
G. Della Valle, M. Ornigotti, E. Cianci, V. Foglietti, P. Laporta, and S. Longhi, Phys. Rev. Lett. {\bf 98}, 263601 (2007).
\bibitem{r35}
M.C Rechtsman, J.M. Zeuner, Y. Plotnik, Y. Lumer, D. Podolsky, F. Dreisow, S. Nolte, M. Segev, and A. Szameit, Nature {\bf 496}, 196 (2013).
\bibitem{r36}
S Longhi, Opt. Lett. {\bf 38}, 3570 (2013).
\bibitem{r37}
I.L. Garanovich, S. Longhi, A.A. Sukhorukov, and Y.S. Kivshar, Phys. Rep. {\bf 518}, 1 (2012).
\bibitem{r38}
T. Meany, M. Gr\"afe, R. Heilmann, A. Perez-Leija, S. Gross, M.J. Steel, M.J. Withford, and A. Szameit, Laser \& Photon. Rev. {\bf 9}, 363 (2015).
 \bibitem{r39}
 L.A. Fernandes, J.R. Grenier, P.R. Herman, J.S. Aitchison, and P.V.S. Marques, Opt. Express {\bf 20}, 24103 (2012). 
 \bibitem{referee1}
S. Longhi, Phys. Rev. Lett. {\bf 101}, 193902 (2008).
\bibitem{referee2}
P.G. Kwiat, K. Mattle, H. Weinfurter, A. Zeilinger, A.V. Sergienko, and Y. Shih, Phys. Rev. Lett. {\bf 75}, 4337 (1995).
\bibitem{referee3}
R.-B. Jin, R. Shimizu, K. Wakui, M. Fujiwara, T. Yamashita, S. Miki, H. Terai, Z. Wang, and M. Sasaki, Opt. Express {\bf 22}, 11498 (2014).
 \bibitem{referee4}
 Y.-H. Li, Z.-Y. Zhou, Z.-H. Xu, L.-X. Xu, B.-S. Shi, and G.-C. Guo,  Phys. Rev. A {\bf 94}, 043810 (2016).
 \bibitem{r41}
  R. Lopes, A. Imanaliev, A. Aspect, M. Cheneau, D. Boiron, and C.I. Westbrook. Nature {\bf 520}, 66 (2015).

 
 \end{thebibliography}
\end{document}